\documentclass[sigconf]{acmart}

\usepackage{multirow}
\usepackage{subfigure}

\AtBeginDocument{%
  \providecommand\BibTeX{{%
    \normalfont B\kern-0.5em{\scshape i\kern-0.25em b}\kern-0.8em\TeX}}}

\copyrightyear{2021}
\acmYear{2021}
\setcopyright{acmlicensed}\acmConference[WI-IAT '21]{IEEE/WIC/ACM International Conference on Web Intelligence}{December 14--17, 2021}{ESSENDON, VIC, Australia}
\acmBooktitle{IEEE/WIC/ACM International Conference on Web Intelligence (WI-IAT '21), December 14--17, 2021, ESSENDON, VIC, Australia}
\acmPrice{15.00}
\acmDOI{10.1145/3486622.3493916}
\acmISBN{978-1-4503-9115-3/21/12}

\begin{document}

\title{Emotion-based Modeling of Mental Disorders on Social Media}

\author{Xiaobo Guo}
\affiliation{%
  \institution{Dartmouth College}
  \city{Hanover}
  \country{USA}}
\email{xiaobo.guo.gr@dartmouth.edu}

\author{Yaojia Sun}
\affiliation{%
  \institution{Dartmouth College}
  \city{Hanover}
  \country{USA}}
  
\email{ys3937@nyu.edu}

\author{Soroush Vosoughi}
\affiliation{%
  \institution{Dartmouth College}
  \city{Hanover}
  \country{USA}}
\email{soroush.vosoughi@dartmouth.edu}

\begin{abstract}
    According to the World Health Organization (WHO), one in four people will be affected by mental disorders at some point in their lives. However, in many parts of the world, patients do not actively seek professional diagnosis because of stigma attached to mental illness, ignorance of mental health and its associated symptoms. In this paper, we propose a model for passively detecting mental disorders using conversations on Reddit. Specifically, we focus on a subset of mental disorders that are characterized by distinct emotional patterns (henceforth called emotional disorders): major depressive, anxiety, and bipolar disorders. Through passive (i.e., unprompted) detection, we can encourage patients to seek diagnosis and treatment for mental disorders. Our proposed model is different from other work in this area in that our model is based entirely on the emotional states, and the transition between these states of users on Reddit, whereas prior work is typically based on content-based representations (e.g., n-grams, language model embeddings, etc). We show that content-based representation is affected by domain and topic bias and thus does not generalize, while our model, on the other hand, suppresses topic-specific information and thus generalizes well across different topics and times. We conduct experiments on our model's ability to detect different emotional disorders and on the generalizability of our model. Our experiments show that while our model performs comparably to content-based models, such as BERT, it generalizes much better across time and topic.
\end{abstract}




\maketitle

\section{Introduction}
   Mental disorders affect a large segment of the public. A report from 2017 estimated that 18.9\% of all U.S. adults have some type of mental health issue \cite{mccance2019national}. The COVID-19 pandemic has most likely increased this number \cite{cao2020psychological}. According to the National Institute of Mental Health (NIMH), the prevalence of anxiety, major depressive, post-traumatic stress, and bipolar disorders among U.S. adults aged 18 or older is much higher than the prevalence of other mental disorders. As stated by the Diagnostic and statistical manual of mental disorders (DSM-5), mood features are the most important and essential features for diagnosing anxiety disorders (AD), major depressive disorder (MDD), and bipolar disorder (BD). According to DSM-5, anxiety disorders are a set of disorders which share features of excessive fear and anxiety; features of major depressive disorder include loss of interest or pleasure in most activities and having consistent depressed moods; bipolar disorder (BD) is a mood disorder characterized by the existence of at least one manic or hypomanic episode and one depressive episode \cite{american2013diagnostic}. 
   Though prevalent in U.S. adults, post-traumatic stress disorder is not as closely related with mood as the other three disorders; it mainly is based on the experience of a shocking, scary, or dangerous event. In this paper, we focus on the three common mental disorders (AD, MDD and BD) which are closely related with mood. Considering their close relationship with emotions, we refer to them as emotional disorders. \footnote{We do not use the term ``mood disorders'' because it is already widely used for naming a set of other mental disorders.}
    
    These disorders can cause great impairment in daily functioning and are often assessed through clinical interviews, brief self-rated and clinician-rated measures.
    Bipolar and related disorders are often related with great impairment in marital and work functioning and increased risk of suicide \cite{maddux2015psychopathology} and Mood Disorder Questionnaire (MDQ) is used for identifying clients. major depressive disorder is associated with impaired cognitive and social functioning \cite{rock2014cognitive,hirschfeld2000social} and Patient Health Questionnaire (PHQ-9) is used for recognizing clients. People diagnosed with anxiety disorders overestimate danger in certain situations and exhibit avoidance behaviors that prevent them from functioning normally and we use Generalized Anxiety Disorder Screener (GAD-7) to figure out clients.
    
    Although these inventories have outstanding screening sensitivity and specificity, they have one significant problem. Because patients must take the initiative and be proactive in participating in completing these inventories, often times potential patients are unwilling to take the survey which will cause undiagnosis or misdiagnosis. Nearly two-thirds of people with a known mental disorder never seek help from a health professional because of stigma, discrimination and neglect \cite{world2001mental}. This problem is complexified by the fact that some emotional disorders will also prevent patients from seeking help. For example, clients with major depressive disorder might not have the energy to proactively seek help, and bipolar disorder is often misdiagnosed for major depressive disorder wrongly because people experience more subjective stress to seek help during the depressive episode \cite{ghaemi2002cade}.
    
    To address this problem, unprompted screening tools have been receiving more attention in recent years, especially those leveraging social media data to gain insight into people's mental states. Most of the existing unprompted screening tools are based on psycholinguistic analysis of the content of user-generated text. For instance, the use of absolutist words \cite{al2018absolute} or the use of first-person pronouns \cite{tackman2019depression,stirman2001word,rude2004language}, have been shown to be predictive of emotional disorders. However, these content-based features capture ``vulnerability factors'' which still exist even after patients have recovered \cite{al2018absolute} and are influenced by topical information \cite{gkotsis2016language}, which raises doubts about the generalizability of the current unprompted screening models and, at best, limit their applications.

    To overcome the shortcomings of unprompted screening models based on \emph{content representations}, we propose an unprompted screening model based on the transition between different emotions expressed by the users on social media. This is inspired by the fact that emotions are topic-agnostic and that different emotional disorders have their own unique patterns of emotional transitions (e.g., rapid mood swings for bipolar disorder, persistent sad mood for major depressive disorder, and excessive fear and anxiety for anxiety disorders). Specifically, we create an emotional ``fingerprint'' for each user by capturing their transition probability matrix of different emotional states. We hypothesize that there will be similarities in the emotional fingerprints between users with similar mental disorders, which can be used for unprompted detection of such disorders. Moreover, our emotion-based model provides greater interpretability, making it more acceptable to mental health clinicians.
    
    Emotions can be manifested in many different modalities such as text, image, audio, and video which makes emotions by nature a multimedia phenomena. In this paper, although we only focus on the emotions represented by text, which is one of the most common representations of emotions on social media, the methods proposed here can be easily extend to other modalities.

    In particular, our work makes the following three contributions:
    \begin{itemize}
        \item We are the first, to the best our knowledge, to leverage emotional transition fingerprints for unprompted models of emotional disorders for users on social media.
        \item We prove that our emotion-based representation (ER) is more robust than content-based representations with our model better generalizing across topics and time.
        \item The results of our analysis of the emotion state transition patterns of users with and without emotional disorders are supported by psychological research on emotional disorders. This is discussed in detail in Section \ref{sec:analysis}
    \end{itemize}
    
\section{Related Work}
    In recent years, social media has become a valuable source for emotional disorders identification and analysis. Several studies have used Twitter data to detect users with major depressive disorder \cite{de2013social}, post-traumatic stress disorder (PTSD) \cite{coppersmith2015clpsych} and bipolar disorder \cite{coppersmith2014quantifying,coppersmith2015adhd,benton2017multi}. Reddit has also been used for studying emotional disorders; specifically for psycholinguistic analysis of emotional disorders \cite{gkotsis2016language,de2014mental,cohan2018smhd,zomick2019linguistic,rude2004language}, the detection of posts indicating anxiety disorders \cite{shen2017detecting} and the detection of users with emotional disorders \cite{sekulic2018not,song2018feature} .
    
    Previous approaches to detect users with emotional disorders have usually relied on the linguistic and stylistic features of user-generated text \cite{coppersmith2014quantifying,sekulic2018not,zomick2019linguistic,sekulic2018not,shen2017detecting}. Inspired by the great success of deep learning methods in the field of natural language processing, deep learning models have also been used for this task. Feature attention networks \cite{song2018feature} and hierarchical attention networks \cite{sekulic2020adapting} which extracts features at post-level and concatenates them at user-level have been built for detecting users with major depressive disorder. These models have shown high interpretability but limited improvement in performance. While concatenating all posts of a user has been shown to be better for this task due to its capture of global features, the interpretability of the model is limited because of its multi-channel design \cite{orabi2018deep}.
    
    However, no existing approaches have focused on the patterns of emotional transition of users, which is not only the core characteristic of emotional disorders, but is also more robust to domain and topic information. In fact, psychologists have demonstrated that the patterns of individuals with emotional disorders are different from those of emotionally healthy individuals regarding emotional reactivity and regulation. For instance, in social anxiety disorder, negative emotions can be detected constantly because of problematic emotional reactivity and deregulation \cite{jazaieri2015role}. At the same time, individuals with major depressive disorder are more likely to be unable to shift from negative emotions to positive emotions compared with healthy individuals \cite{pe2015emotion}.
    
    Built on previous research, we proposal the ER method to represent user posts which is based on the patterns of emotional transition of users, and focus on the global features.

\section{Dataset}
    Reddit is one of the largest social media sites in the world, with over 430 million monthly active Reddit users. It is suitable for our study not only because of its vast number of users, but also because it offers user anonymity and covers a wide range of topics which makes our model more robust across different topics. Moreover, all posts and comments are available publicly from pushshift \cite{baumgartner2020pushshift}, which was used to obtain our dataset.

    We collected two sets of data from 2011 to 2019: 1) A dataset of users with one of the three emotional disorders of interest (BD, MDD, AD); and 2) A dataset of users without known mental disorders serving as the control group. We ensured that our data for the four classes (three emotional disorders plus the control group) had similar temporal distributions, in order to avoid temporal information leakage that could help the model separate the classes. The data we collected was balanced, with 1,997 users for each of the four classes. The data was randomly split into training (70\%), validation (15\%), and test (15\%). Below, we describe how each of the two sets of data were collected. Note that we did not include the data for 2020 because it was very imbalanced (i.e., very few data points with emotional disorders), which would have artificially boosted the performance of the models.
    
    \subsection{Users with Emotional Disorders}
        To obtain a sample of users with emotional disorders, following previous work \cite{beller2014ma,coppersmith2014quantifying}, we used distant supervision and looked for users with self-reported emotional disorders. To do so, we searched user posts (comments and submissions) for the string ``I am diagnosed with bipolar/depression/anxiety" in English and its paraphrased versions (including ``I have been diagnosed", ``I was diagnosed",``I'm diagnosed",``I have been diagnosed", and ``I was just diagnosed") from all subreddits. Though the list is not complete, this can provide us a large sample of users with emotional disorders. We collected all posts of users with self-reported emotional disorders.

         To estimate the labelling quality, we randomly sampled 100 users for each emotional disorder and manually checked their posts. As we found no false positives (i.e., all users reported being diagnosed with the labelled emotional disorder), the dataset is of high precision. However, the true precision of the dataset depends on the veracity of the self-report diagnosis, which we cannot verify.
        
        To make the analysis reliable and unbiased, we decided to additionally prune the dataset as follows. To remove the influence of the self-reports on the users behavior, we only keep the posts before self-reports and remove all users without posts before self-reporting because of the following reasons: First, prior work shows that clients’ realization that they have emotional disorders will influence the behavior and feelings of the clients \cite{farina1971mental,farina1968role}. Second, our work is aimed at detecting users with emotional disorders at the early stages (in order to encourage users to seek further help), thus we want to focus on posts before diagnosis. Since we do not know the actual time of diagnosis, we use self-reports as a proxy. We also exclude all users that report being diagnosed with more than one emotional disorder since we cannot know which emotional disorder, if any, is the dominant disorder, and whether the combination of emotional disorders changes the overall emotion patterns. 
        
    \subsection{Control Group}
        The control group is sampled from the general Reddit community, Although it is possible that some users in the control group have either undiagnosed or unreported mental disorders, their effect would be diluted given the large number of users in our study. 
        
        Additionally, to avoid inadvertently training a topic classification model instead of an emotional disorder detection model, we controlled the topic similarity of different groups. To ensure that the topics discussed by the users in the control group are similar to those talked about by users with emotional disorders, we sample users (who have not self-reported) from the top 20 subreddits  \footnote{TranscribersOfReddit, TalkativePeople, AskReddit, longtail ,Nudelete ,politics, teenagers, PUBGvideos, news, AMAAggregator, relationships, funny, unpopularopinion, worldnews, todayilearned, SquaredCircle, bipolar, AmItheAsshole, pics, nfl} most frequently used by users with emotional disorders. 
        
        Finally, we removed all users from the control group who posted in subreddits about emotional disorders \footnote{mentalhealth, bipolar, bipolar2, BipolarReddit, BipolarSOs, bipolarart, depression, Anxiety, Anxietyhelp, socialanxiety} because these users are likely to be ``false negatives'' (i.e., having an emotional disorder but not self-reporting).
        
        We collected all user posts in the control group. To estimate the labelling quality, we randomly sampled 100 users from the control group and manually check their posts. None of them reported being diagnosed with bipolar, major depressive or anxiety disorders.
        
    \subsection{Time Period Prune}
        By analyzing the temporal distribution of posts of users of the two sets,
        we found control group users are more active in 2020 and the users with emotional disorders are more active before 2011. One reason for the phenomenon is that we only keep the data of users with emotional disorders before self-reporting. 
        
        To eliminate the influence of temporal distribution on content-based representation, we only keep the data between 2011 and 2019 in which the distribution of control group and users with emotional disorders are similar. To build a balanced dataset, we downsample the users in the classes of bipolar, major depressive and control group based on the number of anxiety. 
        
        The final number of each classes (control group, bipolar, major repressive and anxiety) is 1997. The sample size of bipolar is 686,359, of major depressive is 914,082, of anxiety is 686,369, and of control group is 2,516,696.

    \section{Emotion Representation (ER) Method}
       To detect emotional disorders, we present a new method to model users on social media, based on the change in their emotional states. This method, the Emotion Representation (ER), is based on the transition probability matrix of user emotional states.
    
        A given user, $U$, is represented as their temporally sorted posts $U=\{P_1,\cdots, P_m\}$, where $P$ corresponds to a post and $m$ is the total number of posts of $U$. Each post, $P$, is the combination of one or more sentences, $P_i = \{T_{i1},T_{i2},\cdots,T_{in}\}$ where $P_i$ is the i-th post of user $U$ (sorted by time), $T$ corresponds to a sentence, and $n$ is the number of sentences in post $P_i$. Each sentence $T$ is labelled with emotion using emotion classifiers.

        To train emotion classifiers, we used the data from the "SemEval shared task on Affect in Tweets", a collection of tweets annotated for eleven emotions: anger, anticipation, disgust, fear, joy, love, optimism, pessimism, sadness, surprise, and trust \cite{mohammad2018semeval}. From these eleven emotions, we chose joy, sadness, anger, and fear which are basic emotions and closely related with emotional disorders according to Ekman \cite{ekman1999basic}.

        For each emotion, we use the pre-trained BERT (LARGE) and turn it into a binary classifier by adding a softmax layer at the end. The model is then fine-tuned using the data from the SemEval “Affect-In-tweets” task. Since there is no high quality Reddit emotion dataset, we trained and evaluated our BERT emotion classifiers on the aforementioned Twitter emotion dataset. We set the BERT sentence length to be 140. We truncate sentences longer than 140 and use the token ‘$<$PAD$>$’ for padding shorter sentences. We also experimented with leveraging multi-label classifiers to label these four emotions, but the performance was much poorer than having four binary classifiers.  We evaluated our BERT emotion classifiers on the test data provided by "SemEval shared task on Affect in Tweets". Considering that Twitter and Reddit are two different social media platforms with quite distinctive language patterns, we also evaluated the emotion classifiers on Reddit data. We randomly sampled 200 sentences whose length is longer than 5 words from Reddit and labeled the emotions manually. The emotions of these 200 comments were labelled manually by three annotators. We used the data points for which there was consensus amongst all three annotators, which corresponded to around 60\% of the comments.
        
        We remove all interrogative sentences because of their ambiguous emotions. The weighted F1-scores on Twitter and Reddit data are presented in Table \ref{tab:emotion_classifier}. As expected, the performance on Reddit data is weaker than on Twitter data (since the model is trained on Twitter data). However, the performance on Reddit data is still much better than random which means that the model can extract emotional signals from the noise

        \begin{table}[!hbt]
        \centering
        \caption{Weighted F1 scores for emotion classifiers on Twitter and Reddit data.}
        \begin{tabular}{lll}
        \hline
        emotion      & Twitter data & Reddit data  \\
        \hline
        anger        & 0.840 &  0.792 \\
        fear         & 0.853 &  0.655  \\
        joy          & 0.855 &  0.718 \\
        sadness      & 0.784 &  0.692 \\
        \hline
        \end{tabular}
        \label{tab:emotion_classifier}
        \end{table}
        
        For a given sentence $T$, we use the fine-tuned emotion classifiers to detect the emotions present in the sentence. Then, we represent $T$ with a four dimensional binary vector $T=(e_1,e_2,e_3,e_4)$ where $e_1,e_2,e_3$ and $e_4$ represent the detection of anger, fear, joy and sadness respectively. For a post $P$ consisting of $n$ sentences ($P = \{T_1,T_2, \cdots ,T_n\}$), we add up all the emotional vectors of each sentence to produce a single four dimensional emotion representation for the post:
        \begin{equation}
         E_P[j] = \sum_{i=1}^{i=n} T_i[j]  ~~~~~~~ \forall j\in\{1,2,3,4\}
        \end{equation}
        
        To generate the emotional transition matrix, we use a sliding window (set to 0.5 hour with step of size 0.5 hour) to pass through all posts of user $U$. For each window, we add the emotional vectors of all the posts. We set any value greater than 1 to be 1 so that at the end we have a four dimensional binary vector for each time window. We define each combination of the four basic emotions (anger, fear, joy and sadness), represented by a four dimensional binary vector, as a different emotional state. This results in 16 unique emotional states. We also add the state (\emph{no-act}) which represents inactivity by the user in the time window, resulting in a total of 17 different states.
        
        We make a Markov assumption \footnote{This assumption is made for calculation convenience.}, in that we assume that the state of a given time period $t_i$ is only related to its previous time period $t_{i-1}$ where $i>1$. Thus, for each user, we generate a 17*17 transition probability matrix, $\{S_{0,0},S_{0,1},\cdots,S_{j,k},S_{16,16}\}$, where $S_{j,k}$ is the probability of changing from state $j$ to state $k$. For $k \in [0,16]$, $\sum_{k=0}^{16}S_{j,k} = 1$ when state $k$ is transitioned to or $\sum_{k=0}^{16}S_{j,k} = 0$ when state $k$ is never transitioned to. 
        
        Using this method, users in our dataset are represented with an emotion transition probability matrix without content information.

   \section{Emotional Disorder Detection}     
    \subsection{Experimental setup}
        We focus on three subtasks of emotional disorder detection: binary classification tasks for distinguishing users with specific emotional disorders (BD, MDD and AD) from the users in the control group.
        
        For the task of emotional disorder detection, we used the ER features defined in the last section to train three classifiers (support vector machines (\textbf{SVM}), logistic regression (\textbf{LogReg}), and random forest (\textbf{RF})). Technically, any model can be used on top of ER. We tried several ML models and Random Forest was the best-performing one. We didn’t try any deep-learning-based models given the small size of our dataset (< 2000 users per class). We tuned and evaluated our models using the validation and test data respectively.
        
        To test whether the ER features is more suitable for the task of emotional disorder identification than content-based representation in terms of performance and robustness, we compare our model with standard content-based models. Because RF was the best performing classifier for term frequency–inverse document frequency (tf-idf) features in previous work, we choose it as one of the baselines. Next, considering that BERT is current state-of-the-art for many natural language processing tasks and since we use BERT for emotional feature extraction, we include a BERT-based emotional disorder classifier in our baselines.
        
        For the tf-idf feature extractor, we concatenate all posts of a user into one document and use the words only appearing in the training data to build the dictionary. To limit the influence of information leakage caused by mental health and medical terminology that is unlikely to be used by the control group users, we remove terms closely related with emotional disorders\footnote{``bipolar", ``anxiety", ``manic", ``depression", ``manic", ``diagnose", ``hypomania", ``pdoc", ``psychiatrist", ``ii", ``therapist", ``mental", ``mood", and ``disorder"} as well as the names of drugs used to treat emotional disorders\footnote{``seroquel", ``lithium"",``lamictal",  ``depakote", ``SSRI", and ``zoloft"}. These terms are chosen because the tf-idf value of the terms in the treatment data are significant higher than the control data, and they are closely related with emotional disorders. We acknowledge that this might not be a complete list, however, adding more terms into the list will only further decrease the performance of tf-idf method, meaning that if anything, we may be overestimating the performance of the tf-idf model.
        
        For the BERT classifier, we encode the posts at the sentence level and use the arithmetic mean of the encoding results as the representation of the user. We use the ``BERT-Large, Cased'' model which has 24 layers, 1024 hidden units, and 16 heads for each layer. We set the BERT sentence length to be 140. We truncate sentences longer than 140 and use the token ‘$<$PAD$>$’ for padding shorter sentences. The output of the 22nd layer is used as the encoding result, as it performed best on the validation data.

    \subsection{Results}
        \subsubsection{ER with different classifiers}
            We fine-tuned each classifier with train and validation data and tested on the test data. As shown in Table \ref{tab:performance}, all three classifiers (SVM, LogReg, and RF) achieve reasonable performance with ER features as measured by four metrics (accuracy, F1-score, precision, and recall). The RF classifier achieves the best performance across different metrics and is thus chosen as our model for the remainder of this paper. Since our dataset is balanced (1,997 data points for each class), the random baseline for the prediction task in Table \ref{tab:performance} is 0.5 for accuracy and other metrics.
            
        \begin{table}[!hbt]
            \centering
            \caption{Comparison between different classifiers based on ER.}
            \begin{tabular}{ccllll}
            \hline
            Metrics                   & Classifier   & \multicolumn{1}{l}{BD} & \multicolumn{1}{l}{MDD} & \multicolumn{1}{l}{AD} \\
            \hline
            \multirow{3}{*}{Accuracy}  & SVM    & 0.791          & 0.771          & 0.782          \\
                                       & LogReg & 0.795          & 0.774          & 0.791          \\
                                       & RF     & \textbf{0.851} & \textbf{0.807} & \textbf{0.839} \\ \hline
            \multirow{3}{*}{F1-score}  & SVM    & 0.790          & 0.771          & 0.781          \\
                                       & LogReg & 0.794          & 0.773          & 0.791          \\
                                       & RF     & \textbf{0.850} & \textbf{0.806} & \textbf{0.838} \\ \hline
            \multirow{3}{*}{Precision} & SVM    & 0.791          & 0.771          & 0.782          \\
                                       & LogReg & 0.795          & 0.774          & 0.791          \\
                                       & RF     & \textbf{0.851} & \textbf{0.807} & \textbf{0.839} \\ \hline
            \multirow{3}{*}{Recall}    & SVM    & 0.795          & 0.775          & 0.789          \\
                                       & LogReg & 0.800          & 0.780          & 0.795          \\
                                       & RF     & \textbf{0.855} & \textbf{0.817} & \textbf{0.844} \\
            \hline
            \end{tabular}
            \label{tab:performance}
        \end{table}

        \subsubsection{Comparison with Baseline Models}   
            
            To compare the performance of ER features and content-based features (i.e., tf-idf and BERT), we fine-tune the hyper-parameters for each model and run 5-fold cross validation to measure the performance of our models. The results are shown in Table \ref{tab:tfidf_and_emotion}.
            \begin{table}[!hbt]
            \caption{Comparison between ER (ours) and baselines using different metrics. Here we show the mean and the standard deviations using 5-fold cross validation.  }
            \centering
            \begin{tabular}{cclll}
            \hline
            \multicolumn{1}{l}{Metrics} & Model  & \multicolumn{1}{l}{BD} & \multicolumn{1}{l}{MDD} & \multicolumn{1}{l}{AD} \\ \hline
            \multirow{3}{*}{Accuracy}   & tf-idf & 0.832(0.005)           & 0.774(0.007)            & 0.806(0.010)           \\
                                        & BERT   & \textbf{0.866(0.005)}  & 0.825(0.007)            & 0.841(0.010)           \\
                                        & ER     & \textbf{0.866(0.017)}  & \textbf{0.832(0.013)}   & \textbf{0.853(0.005)}  \\ \hline
            \multirow{3}{*}{F1-score}   & tf-idf & 0.832(0.005)           & 0.774(0.007)            & 0.806(0.010)           \\
                                        & BERT   & \textbf{0.866(0.005)}  & 0.825(0.007)            & 0.841(0.010)           \\
                                        & ER     & 0.865(0.017)           & \textbf{0.831(0.013)}  & \textbf{0.852(0.005)}  \\ \hline
            \multirow{3}{*}{Recall}     & tf-idf & 0.832(0.005)           & 0.774(0.007)            & 0.806(0.010)           \\
                                        & BERT   & \textbf{0.866(0.005)}  & 0.825(0.007)            & 0.841(0.010)           \\
                                        & ER     & \textbf{0.866(0.017)}           & \textbf{0.832(0.013)}   & \textbf{0.853(0.005)}  \\ \hline
            \multirow{3}{*}{Precision}  & tf-idf & 0.832(0.005)           & 0.774(0.007)            & 0.807(0.010)           \\
                                        & BERT   & 0.866(0.005)           & 0.826(0.007)            & 0.841(0.010)           \\
                                        & ER     & \textbf{0.869(0.016)}  & \textbf{0.839(0.012)}   & \textbf{0.857(0.004)}   \\ \hline
            \end{tabular}
            \label{tab:tfidf_and_emotion}
            \end{table}
            Our ER model (random forest model trained with ER features) achieves better performance than the tf-idf model in all three tasks. Our model's performance is similar to the performance of the BERT model in the difference tasks. Our model performs slightly better in the anxiety and major depressive tasks and slightly worse in the bipolar task. This phenomenon might suggest that these two ways of modeling users are complementary. Since the focus of this work is the introduction of emotion representation and comparison to content-based representations, we do not explore models that combined emotion and content representations, though this is an exciting area for future research.
            
            Considering that the performance between BERT and our ER features are similar, we conduct Welch's t-test on the performance of BERT and ER features with the null hypothesis that the performances of these two models are same. The p-values, shown in Table \ref{tab:p_baseline}, show that the difference between the BERT and the ER models are not significant (with the possible exception of AD), meaning that they achieve similar performance on all tasks across all metrics (while ER's advantage over td-idf is shown to be  statistically significant).
            \begin{table}[!hbt]
            \centering
            \caption{p-values for the null hypothesis that the performances of ER and BERT/tf-idf are the same (using a t-test). Acc=accuracy, F1=F1-score, Re=recall and Pre=precision. * corresponds to p<0.05.}
            \begin{tabular}{clll|clll}
            \hline
            \multicolumn{4}{l|}{ER vs. BERT}                                                              & \multicolumn{4}{l}{ER vs. tf-idf}                                                           \\ \hline
             & \multicolumn{1}{l}{BD} & \multicolumn{1}{l}{MDD} & \multicolumn{1}{l|}{AD} &   & \multicolumn{1}{l}{BD} & \multicolumn{1}{l}{MDD} & \multicolumn{1}{l}{AD} \\ \hline
            Acc  & 0.978                   & 0.411                    & 0.077                    & Acc  & 0.011*                   & 0.000*                    & 0.000*                   \\
            F1  & 0.955                   & 0.486                    & 0.088                    & F1  & 0.012*                   & 0.000*                    & 0.000*                   \\
            Re    & 0.978                   & 0.412                    & 0.077                    & Re    & 0.011*                   & 0.000*                    & 0.000*                   \\
            Pre & 0.770                   & 0.086                    & 0.025*                    & Pre & 0.006*                   & 0.000*                    & 0.000*                   \\ \hline
            \end{tabular}
            \label{tab:p_baseline}
            \end{table}

            It is important to ensure all users with emotional disorders are identified, while at the same time not misidentify users without emotional disorders. To do so, we compare the false positive rate of the ER model and baseline models when the true positive rate is 1 (i.e., all users who have an emotional disorder have been identified). Again, as shown in Table \ref{tab:false_positive_rate}, our model outperforms the tf-idf model and does as well as the BERT model, though on different tasks. Specifically, on the task of bipolar detection, ER achieves the best (lowest) false positive rate when the true positive rate is 1, while BERT performs best on the major depressive and anxiety tasks.
            
            \begin{table}[!hbt]
            \caption{The false positive rate of ER and baselines when true positive rate is 1.}
            \centering
            \begin{tabular}{cllll}
            \hline
            \multicolumn{1}{l}{Model} & BD        & MDD    & AD        \\
            \hline
            tf-idf                    & 0.833          & 0.983          & 0.880           \\
            BERT                      & 0.863          & \textbf{0.779} & \textbf{0.846} \\
            ER                        & \textbf{0.649} & 0.953          & 0.943          \\
            \hline
            \end{tabular}
            \label{tab:false_positive_rate}
            \end{table}

            From these experiments we can conclude that the ER model is more suitable for the task of emotional disorder prediction than tf-idf, while the performance of ER and BERT are similar (though seemingly complementary) to each other for this task.

            \begin{figure*}[t]
                    \centering
                    \subfigure[BD]{
                    \includegraphics[width=0.63\columnwidth]{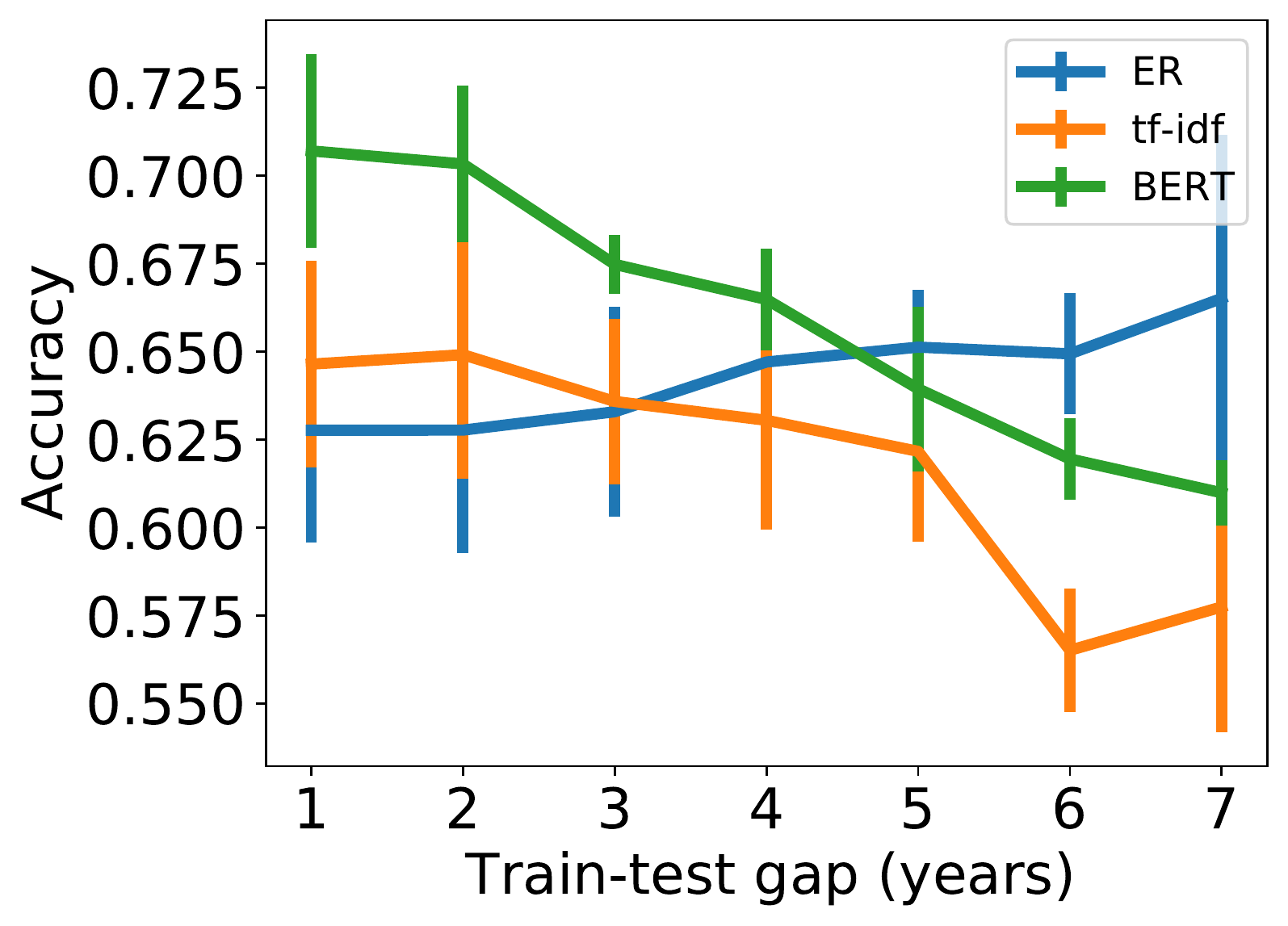}
                    }
                    \subfigure[MDD]{
                    \includegraphics[width=0.63\columnwidth]{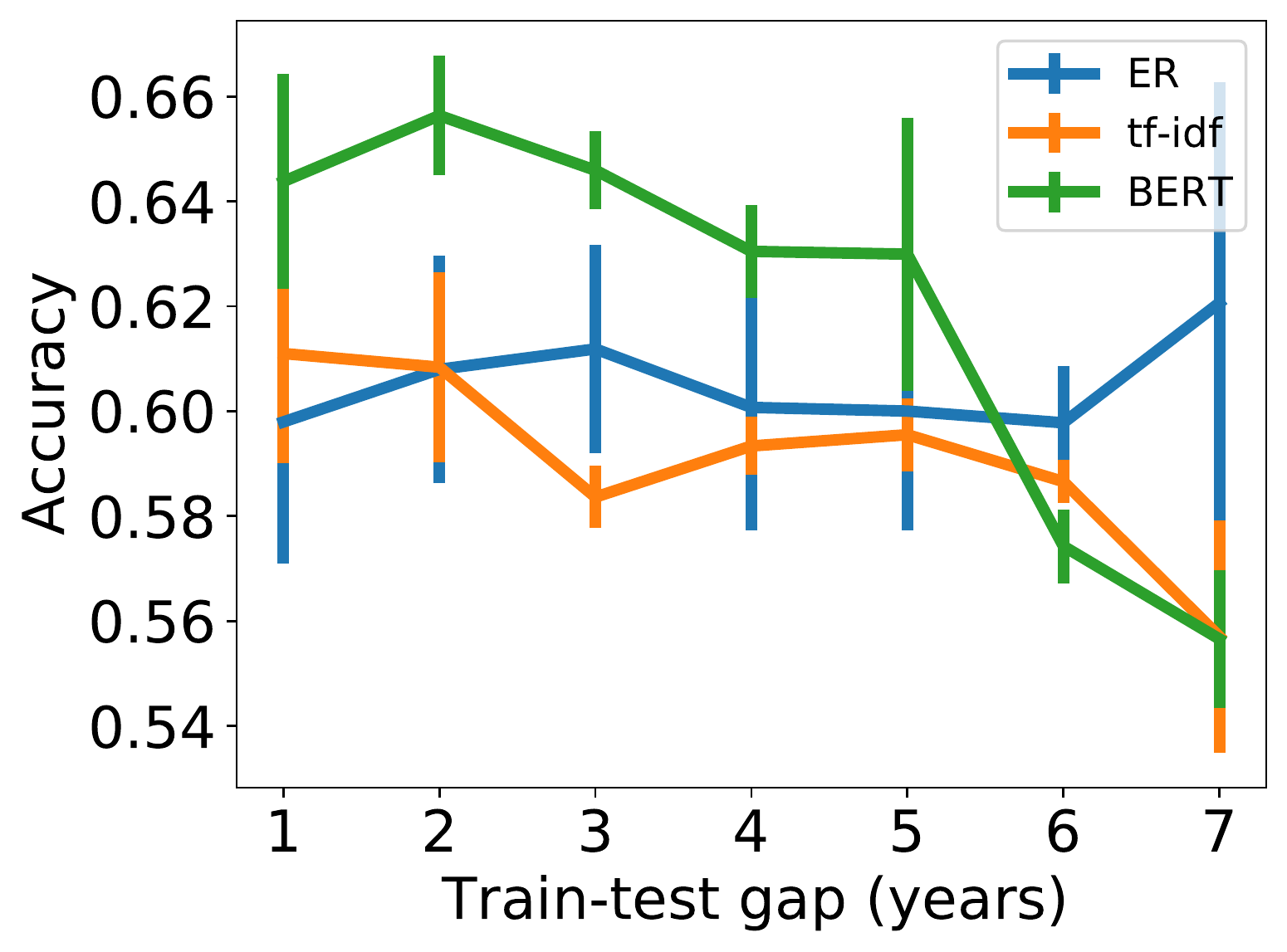}
                    }
                    \subfigure[AD]{
                    \includegraphics[width=0.63\columnwidth]{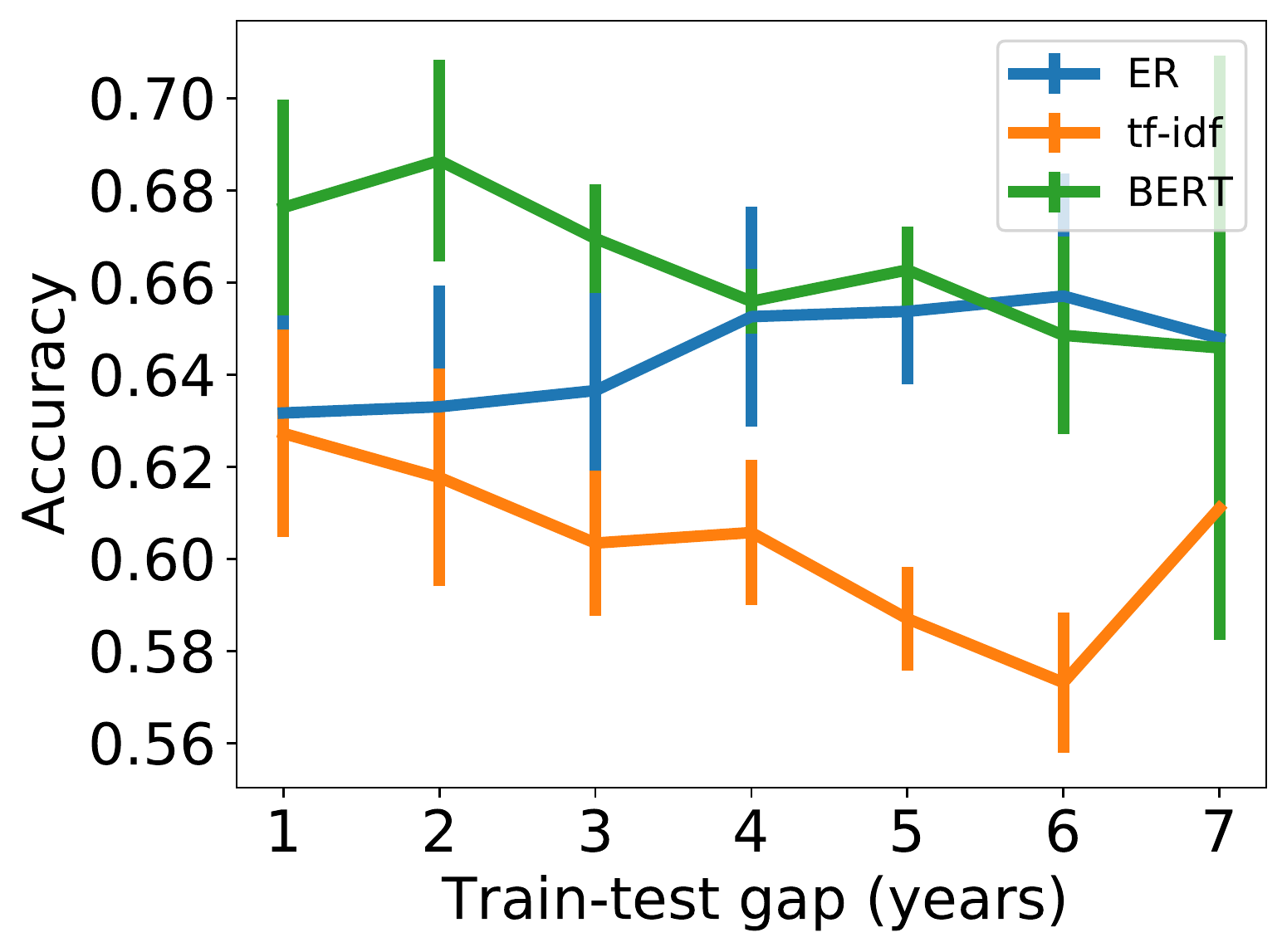}
                    }
                \caption{The mean and standard error of the accuracy for ER and baselines with train and test data on different years}
                \label{fig:gap}
            \end{figure*}       
 
        \subsubsection{Generalizability}
            
            In this paper we define generalizability as the robustness of a trained model across time (since topics of conversations are prone to change over time). To compare the generalizability of the ER and the baseline models, we use \emph{temporal testing}, which entails training and testing a model on data from different years. If a model captures topical or domain information, then its performance should decrease as the gap increases. Conversely, if a model accurately captures the underlying signals of emotional disorder, then its performance should be stable.
            
            Temporal testing is conducted for different gap sizes (from 1 to 7 years) between train and test data. As the gap increases, the number of experiments conducted decreases. This is because with a gap size of 1, experiments are conducted on every consecutive year, while for a gap size of 7, test data should be 7 years apart from the training data. For each gap year, we calculate mean and standard error of the results for all experiments with balanced train and test data. We ensure that there are no duplicate users in train and test data.

            As show in Figure \ref{fig:gap}, the performance of the ER model is the most stable compared to the baseline models. The performance of BERT and tf-idf models decrease as the gap between training and test data increases (sometimes drastically as indicated by Fig \ref{fig:gap} (c)), while the ER model maintains its high-level performance.
            % even after 7 years.
            The exact difference is shown in Table \ref{tab:acc_gap} where negative means the performance of seven years gap is worse than that of one year gap , and vice versa.
            
            \begin{table}[!hbt]
            \caption{Difference in accuracy between the 7 year and 1 year train-test gaps. }
                \begin{tabular}{clll}
                \hline
                      & tf-idf & BERT   & ER(ours) \\
                \hline
                BD    & -0.069 & -0.097 & 0.037    \\
                MDD   & -0.054 & -0.087 & 0.023    \\
                AD    & -0.016 & -0.030 & 0.016    \\
                \hline
                \end{tabular}
                \label{tab:acc_gap}
            \end{table}
            
            To figure out whether the differences between the performances of the models with one year and seven years gaps are significant or not, we conduct Welch's t-test on the performance of one year and seven years gaps. The p-values for the null hypothesis that the performances of one year and seven years gaps are the same are shown in Table \ref{tab:p-value_gap}. We can observe that on all tasks, the performances of the ER method are stable since the differences are not significant. Meanwhile, for the BERT method, the decreases between the one year and seven years gaps are significant for the BD and MDD tasks. Though the decreases for the tf-idf model are not significant, the model generally performs much worse than the ER model, as shown in Table \ref{tab:tfidf_and_emotion}. This temporal testing demonstrates that our model indeed captures the underlying signals of emotional disorder, while the content-based models, BERT and tf-idf, seem to, at least to some level, capture time-specific signals. However, we also notice that the standard errors of our experiments are also large, which might be caused by the limited number of data points for each gap year.

            \begin{table}[!hbt]
                \caption{p-values for the null hypothesis that the mean performances of one year and seven years gaps are the same (using a t-test). * corresponds to p<0.05.}
                \begin{tabular}{clll}
                \hline
                      & tf-idf & BERT  & ER(ours) \\
                \hline
                BD    & 0.243  & 0.010*  & 0.572    \\
                MDD   & 0.169  & 0.011* & 0.696    \\
                AD    & 0.509  & 0.716 & 0.804    \\
                \hline
                \end{tabular}
                \label{tab:p-value_gap}
            \end{table}
            
            These results, combined with the results from the previous experiments where our model performs better than tf-idf and comparably to BERT, show that our ER model is the most robust for emotional disorder detection on social media.

\section{Discussion}
    Here, we further conduct a set of analyses to better understand what our emotion representations capture by studying the difference between emotion transition patterns of users with and without disorders and by probing the generalizability of our emotion representations. 
    
    \subsection{Emotion Analysis}
        \label{sec:analysis}
         As atypical (and problematic) patterns of emotional reactivity and regulation are the main symptoms of emotional disorders, here we explore observed patterns of emotional state transition among users of different classes (i.e. control group, bipolar, major depressive, and anxiety users) and discuss how well our observed patterns align with prior research on emotional disorders.
        To conduct this analysis, we compare the transition matrix of emotion states for different disorders. Because the probability of changing to no-act state (meaning inactivity on Reddit) is dominant for each emotion state in the transition matrix, we remove the no-act state for better representation. We calculate the transition probability matrix of one class of users by averaging all transition matrices of users from the same class. 
            
        \begin{figure*}[!hbt]
            \centering
                \subfigure[Control Group]{
                \includegraphics[width=0.49\columnwidth]{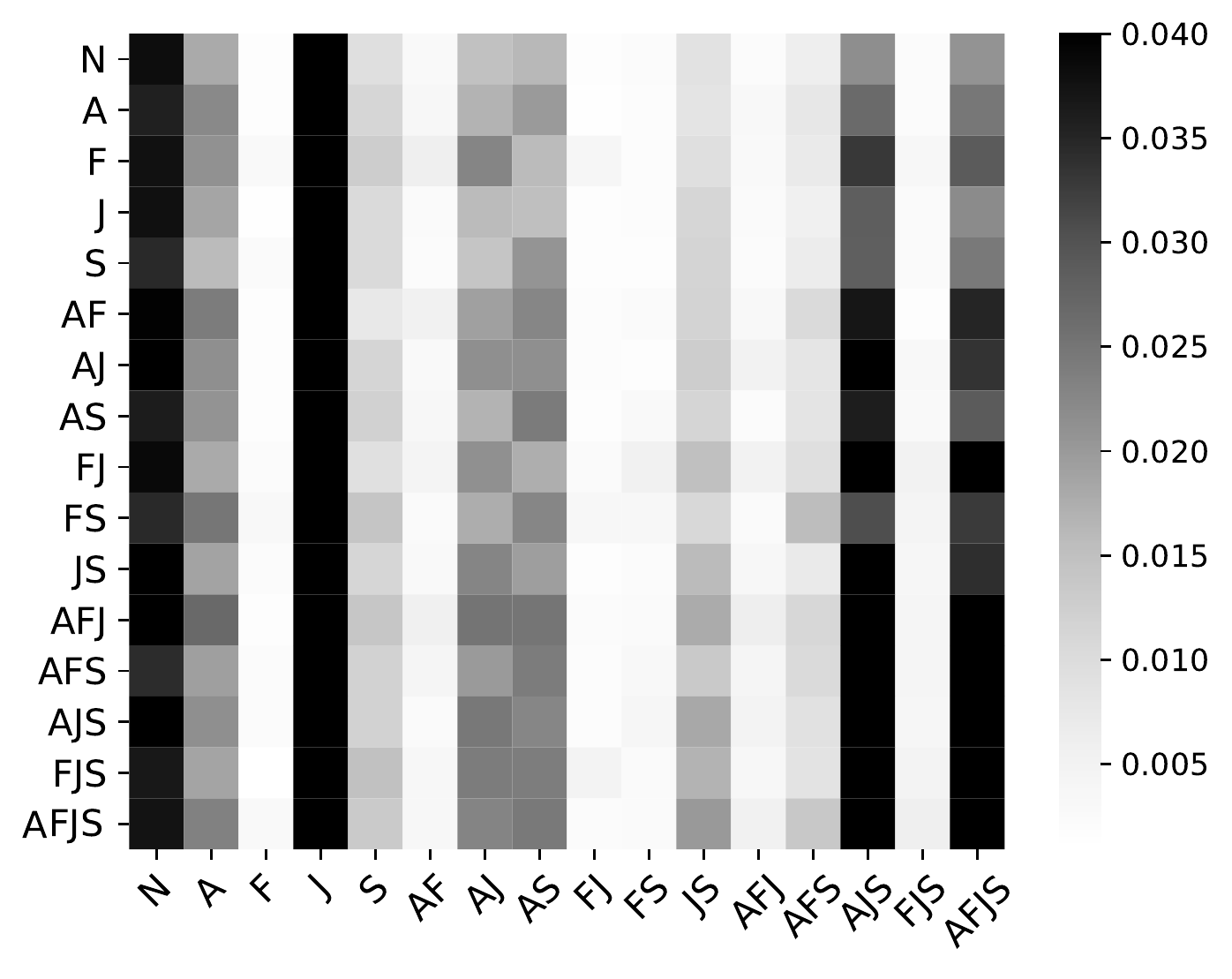}
                }
                \subfigure[BD]{
                \includegraphics[width=0.49\columnwidth]{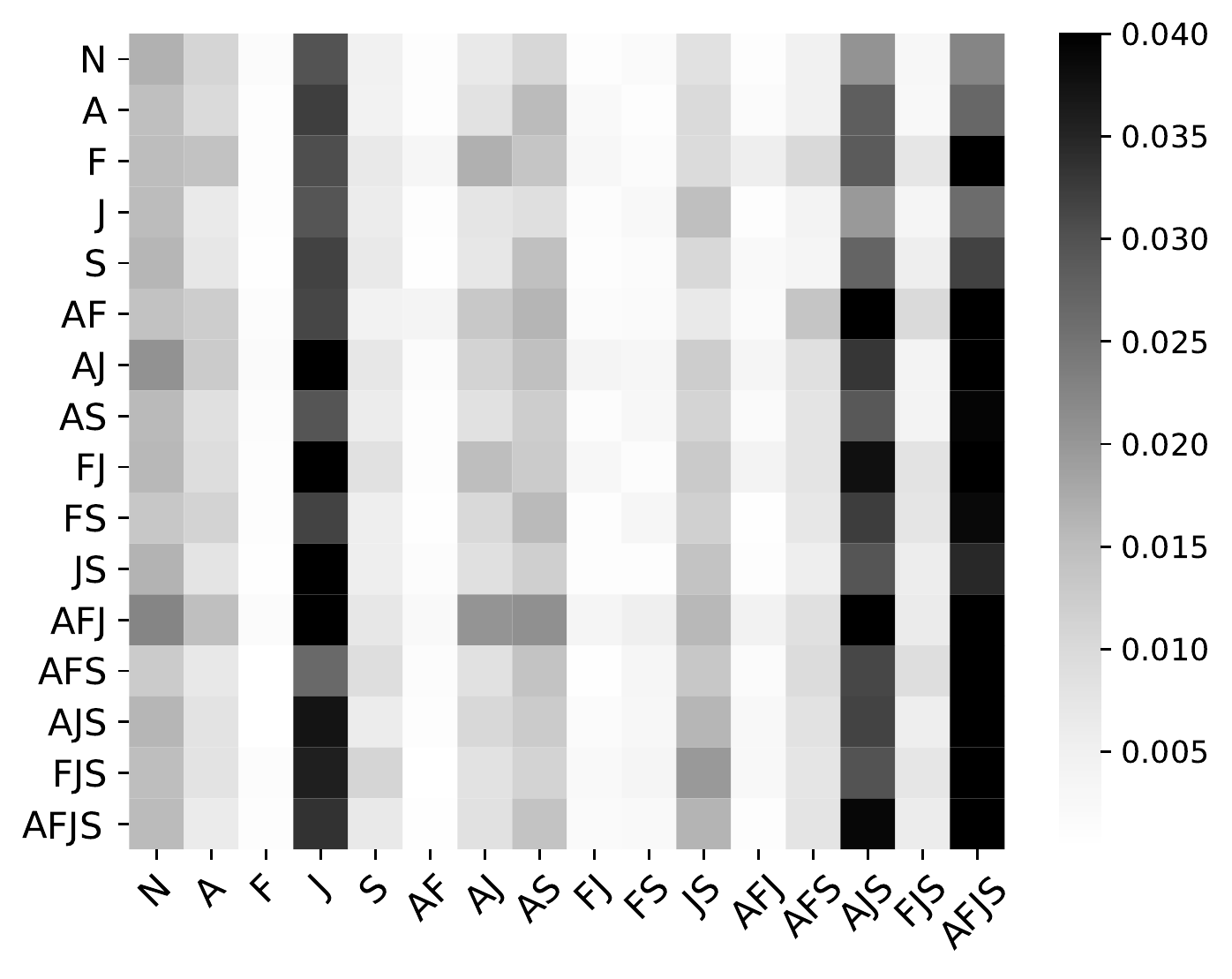}
                }
                \subfigure[MDD]{
                \includegraphics[width=0.49\columnwidth]{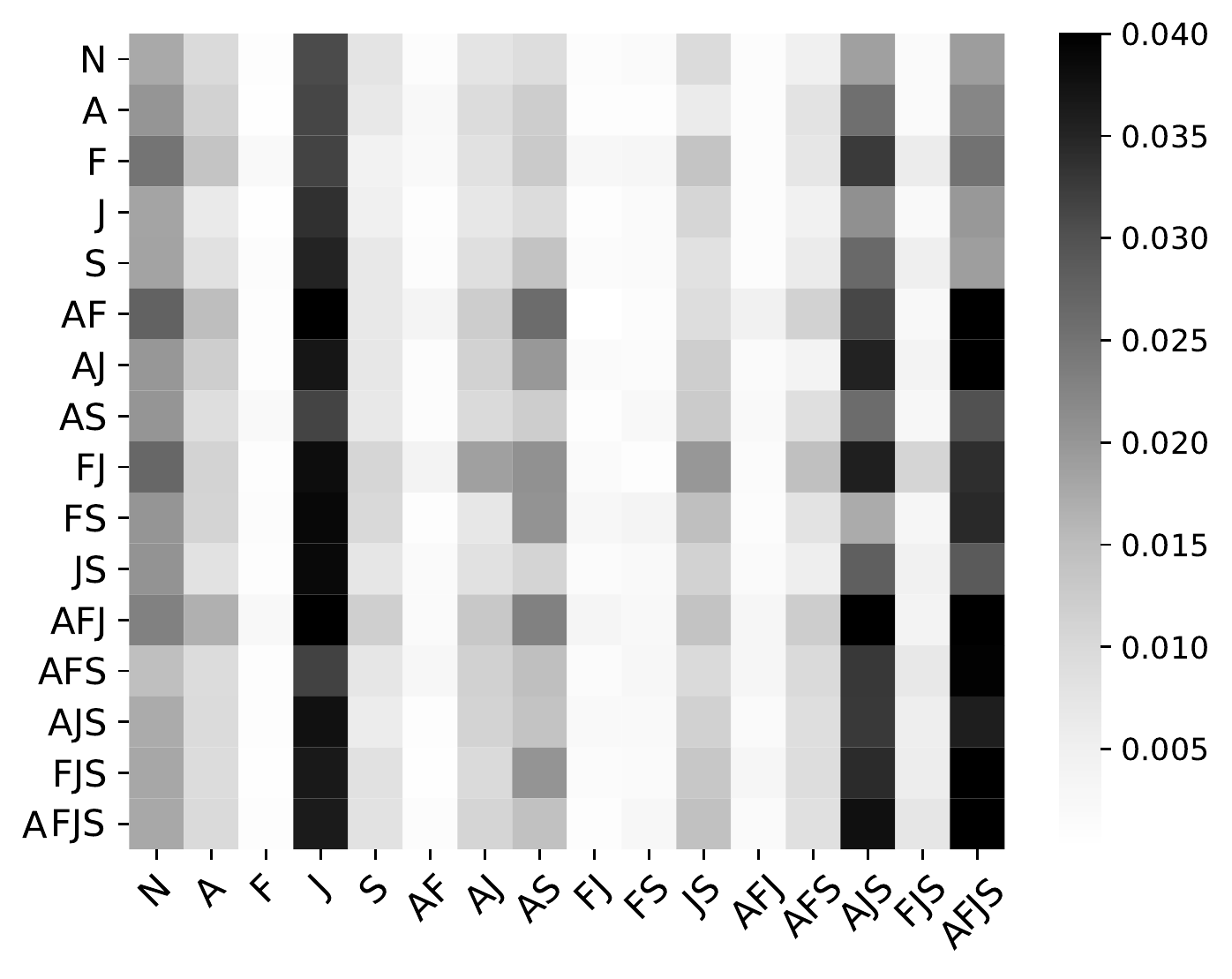}
                }
                \subfigure[AD]{
                \includegraphics[width=0.49\columnwidth]{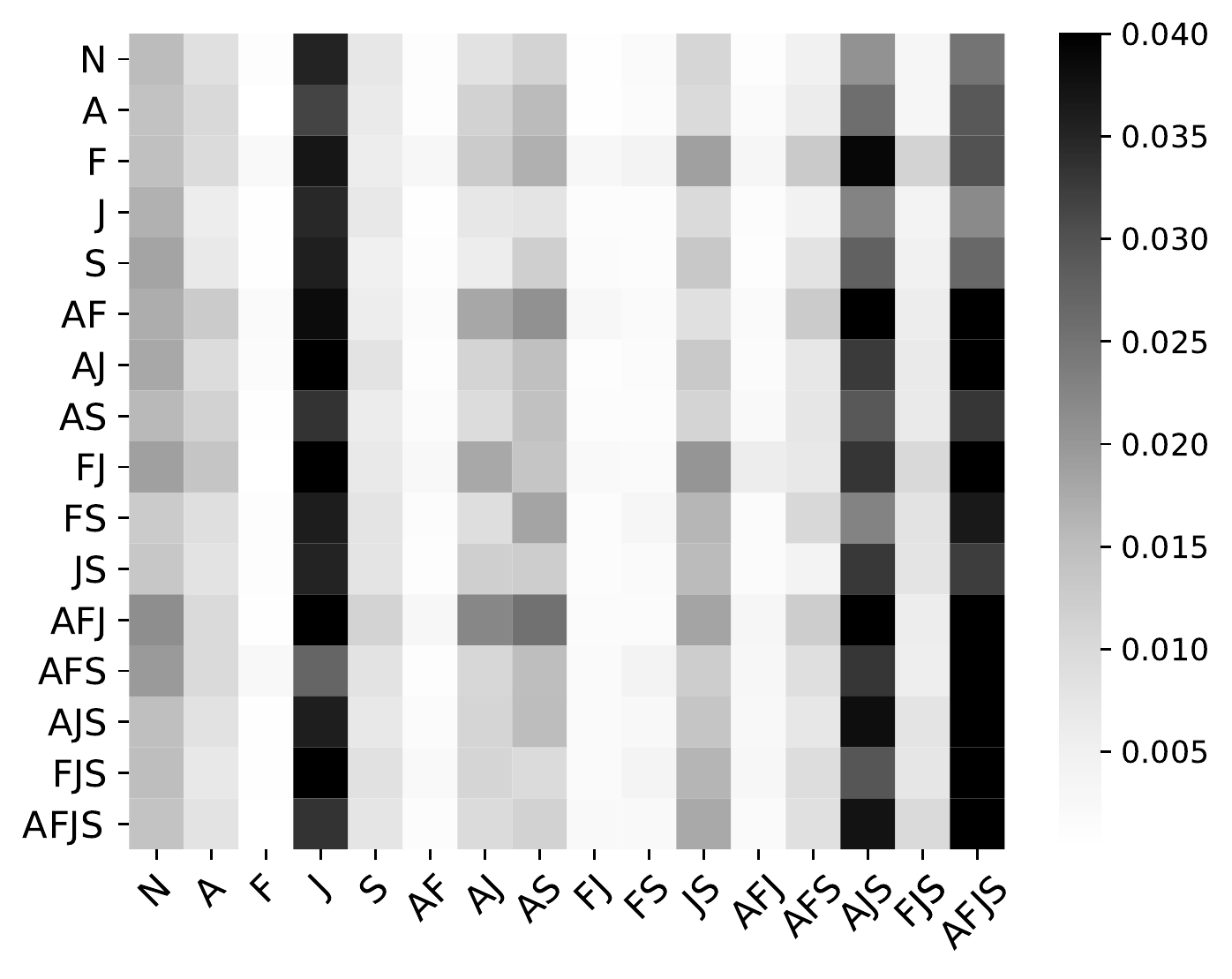}
                }
            \caption{Transition matrix of emotion states for emotionally healthy users (control group), user with bipolar disorder (BD), users with major depressive disorder (MD) and users with anxiety disorders (AD). N: no emotion, A: anger, F: fear, J: joy, S: Sadness. The states which include several emotions are represented with multi characters (e.g., 'AF' means the combination of anger and fear).}
            \label{fig:transition_matrix_state}
        \end{figure*}
            
             As shown in Figure \ref{fig:transition_matrix_state}, we generate the transition matrix of emotion states for emotionally healthy users (Figure \ref{fig:transition_matrix_state} (a)), users with bipolar disorder (Figure \ref{fig:transition_matrix_state} (b)), users with major depressive disorder (Figure \ref{fig:transition_matrix_state} (c)), and users with anxiety disorders (Figure \ref{fig:transition_matrix_state} (d)). The rows indicate the current emotional state, and the columns indicate the target emotional state.
             N represents no emotion; 'A', anger; 'F', fear; 'J', joy; and 'S', sadness. The states which include several emotions are represented with multiply characters (e.g., 'AF' means the combination of anger and fear). Here, the rows represent the original states and the columns represent the target states.
            
            It's evident that the transition matrix of control group users is different from users with emotional disorders, in particular column 'N', and certain columns containing 'A' (states A, AJ, AS, AJS, and AFJS). These two significant differences are in accordance with the symptoms of emotional disorders described earlier in the paper.
            
            From Figure \ref{fig:transition_matrix_state}, we observe that emotionally healthy (i.e. control group) users are more likely to return to states without any discernible dominating emotion (state N). This phenomenon points to a shared feature among emotional disorders, which is impaired emotional regulation ability to ``monitor, evaluate and modify emotional reactions'' \cite{thompson1991emotional} and recover quickly from high arousal states. Similarly, the lower probability of changing to the joy state (state J) in Figure \ref{fig:transition_matrix_state} (b), (c), and (d) compared with Figure \ref{fig:transition_matrix_state} (a) aligns with research showing that users with emotional disorders tend to suffer from emotional disturbances (i.e. a general pervasive mood of unhappiness or depression). 
            
            The results also show that emotionally healthy users are more likely to present with states that involve anger (i.e. states A, AJ, AS, AJS, and AFJS), as can be seen in Figure \ref{fig:transition_matrix_state} (a) compared with Figure \ref{fig:transition_matrix_state} (b), (c), and (d). Anger is often triggered by stimuli that threaten our self-esteem. Research shows that people with discrepant self‐esteem are more likely to exhibit greater anger suppression, a more depressive attributional style, and more nervousness \cite{schroder2007high}. When facing self-esteem threatening stimuli, people with emotional disorders tend to suppress anger and instead transition into sadness.
            
            Additionally, we speculate that the high correlation between the control and disorder groups shows that emotions exhibited in posts is affected by many factors other than emotional disorders (e.g., one would expect the emotion of users to transition to be sadder because of the pandemic). Also, users that have disorders with extreme symptoms are probably less likely to be active which makes the difference between the groups appear milder than it may be in reality.

    \subsection{Generalizability}
        While the generalizability of any predictive model is important, this importance is magnified for models used for medical diagnosis or intervention. A model which is sensitive to the change of time periods is more likely to be sensitive to the change of topics, language habits and other factors. A model with low specificity will not only increase the work of psychotherapists dramatically due to misdiagnosis, but also negatively influence the emotionally healthy users who are wrongly labelled as having emotional disorders. Additionally, a model which is not robust over long time periods needs to be retrained frequently which requires regular and time-consuming data collection and annotation expensive. In this section, we discuss different factors that could affect generalizability and how our model addresses these problems. 
        
        One well known factor that has a clear influence on generalizability is the difference in topic or domain of the training and test data. While content-based representations like BERT are powerful, they inadvertently capture irrelevant topic or domain information that will artificially boost performance at the cost of over-fitting and reduction in generalizability. Our emotion representations suffer less from this problem. Although the emotions are detected using content, the abstraction of emotion representations dilute any topic-specific information. As shown in Figure \ref{fig:gap}, our emotional representation is more robust and generalizable than content-based models.
        
        The second factor that affects the generalizability of content-based representation is information leakage, which in our case refers to the phenomenon that users with emotional disorders talked about the emotional disorders, symptoms, or treatment before self-reporting. Through examining the tf-idf model, we observed that certain words that are closely related with emotional disorder symptoms (e.g., ``hypomanic'' and ``episode'') and treatment (e.g., ``seroquel'' and ``lithium'') are of great importance in prediction. When applying this type of content-based model on users without any diagnosis, knowledge of treatments, or names of symptoms, the model's performance will be reduced and the number of cases of misclassification will increase. However, because our representation does not contain words related with emotional disorders, models based on our emotion representation do not have to be concerned about this problem.
        
        The third problem which affects the generalizability of models relying on content-based representation is irrelevant features. These models are likely to learn features irrelevant or even unwanted to the task at hand, and what features are learnt cannot be controlled. For example, we noticed that the temporal distribution of our training data is different for different classes. Even though the number of users for each class is the same, the users in the control group class post more after 2018 compared to the users in the three emotional disorder classes. This means that the prior probability for a randomly chosen user who only posts after 2018 is weighted toward the control group. Therefore, although time should not be used as a feature for our task (even if it is predictive), the predictor based on features extracted by tf-idf recognizes some key words such as "RTX 2080ti" (the name of a graphics card which was released in 2018) and uses them as a proxy for time information to make a prediction. This again will artificially inflate the evaluation performance of these content-based models at the cost of damaging their generalizability. Models based on our emotion representation can avoid learning irrelevant or unwanted features.
        
        Therefore, compared with content-based representation, our emotion representation has better generalizability for emotional disorder prediction task because it dilutes the influence of topic change, and avoids the problems of information leakage and irrelevant features.

\section{Conclusion \& Future Work}
     To the best of our knowledge, this paper is the first to leverage emotion states for identifying users with emotional disorders (bipolar, depressive, and anxiety disorders) on social media. For this task, we propose a topic-agnostic method based on an emotional transition probability matrix generated by the emotion states in user-generated text. We find that a simple random forest classifier trained on a 17x17 emotional transition matrix can outperform a more complex tf-idf based classifier and perform comparably to a BERT classifier.
    
    More importantly, our approach, different from content-based representations influenced by topic, domain, and information leakage, is more robust and has better interpretability.
    
    In our future endeavors, we plan to further validate our emotion representations by exploring a wider range of classifiers. Moreover, to test whether our method is platform independent, we plan to extend our approach to other social media platforms, such as Twitter. Additionally, our current approach of using an emotion classifier to characterize the emotional states of users can be improved by either having more accurate emotion classifiers or by exploring new methods to characterize the emotional states of users. We also plan to study whether other non-content features (e.g. post time, post count, etc.) will be helpful for identifying users with emotional disorders. Finally, a future avenue for research is to integrate other modality such as image, audio, and video into our model to strengthen its performance, generalizability, and interpretability. 

\bibliographystyle{ACM-Reference-Format}
\bibliography{main}
\end{document}